\documentclass[10pt,conference]{IEEEtran}

\usepackage[T1]{fontenc}
\usepackage[utf8]{inputenc}
\usepackage{balance}
\usepackage{booktabs}
\usepackage{graphicx}
\usepackage{listings}
\usepackage{tikz}
\usetikzlibrary{positioning,fit,calc,arrows.meta}
\usepackage{url}
\usepackage{xcolor}
\usepackage{cite}
\usepackage{orcidlink}

\newcommand{\sys}{\textsc{Stratum}}

\newcommand{\evSucA}{6/24}   
\newcommand{\evSucB}{21/24}  
\newcommand{\evCallsA}{18}   
\newcommand{\evCallsB}{9}
\newcommand{\evGetsA}{6.5}   
\newcommand{\evGetsB}{5}
\newcommand{\evTokA}{196.6}  
\newcommand{\evTokB}{43.3}

\lstdefinelanguage{json}{
  basicstyle=\ttfamily\scriptsize,
  showstringspaces=false,
  breaklines=true,
  columns=fullflexible,
  keepspaces=true,
}

\begin{document}

\title{\sys{}: Repository Memory for Coding Agents\\Beyond the Context Window}

\author{
\IEEEauthorblockN{Yuvraj Sehgal\,\orcidlink{0009-0002-1980-604X}}
\IEEEauthorblockA{\textit{Department of Computer Science}\\
\textit{Brock University}\\
St.\ Catharines, Ontario, Canada\\
ys19rk@brocku.ca}
\and
\IEEEauthorblockN{Mahsa Panahandeh\,\orcidlink{0000-0002-6369-8982}}
\IEEEauthorblockA{\textit{School of Electrical Engineering}\\
\textit{and Computer Science}\\
\textit{University of Ottawa}\\
Ottawa, Ontario, Canada\\
mpanahan@uottawa.ca}
\and
\IEEEauthorblockN{Naser Ezzati-Jivan\,\orcidlink{0000-0003-1435-6297}}
\IEEEauthorblockA{\textit{Department of Computer Science}\\
\textit{Brock University}\\
St.\ Catharines, Ontario, Canada\\
nezzatijivan@brocku.ca}
}

\maketitle

\begin{abstract}
Context windows have grown a thousand-fold, yet coding agents still start every session with no memory of their past work. They re-read files, retry failed approaches, and forget root causes they established hours earlier. This is not a context-length problem but a memory-architecture problem. Our hypothesis: \emph{coding agents need persistent, experience-aware repository memory, so that knowledge survives from one session to the next.} Three measurements on real repositories motivate the design. First, mid-size projects already exceed the largest production context windows, and storing knowledge inside the model's key--value (KV) cache costs $10^4$ to $10^5$ times more bytes than storing it as indexed text. Second, across 752 real bug-fix commits, the median fix edits two files, and these files plus their direct imports amount to less than 10\% of the repository; the bottleneck is selection, not capacity. Third, even with transitive import reachability and test links, 47\% of co-edited file pairs remain structurally disconnected; the knowledge needed for selection must be mined from history and experience. We sketch \sys{}, a tiered memory that lives with the repository and serves it through page-safe tools: every response states what was returned, what remains, and what it cost. Retrieval over a store of one million nodes takes 1.7\,ms at the median. In a multi-session experiment on eight real bugs, a fresh agent that inherited a prior session's memory, written by a different model, identified the root-cause file in 21 of 24 trials versus 6 of 24 for a retrieval-only baseline (Fisher $p{<}10^{-4}$), at $4.5\times$ lower median token cost. With the notes of all eight investigations pooled in one store, success remained 8/8.
\end{abstract}


\section{Introduction}
Autonomous coding agents now routinely work for hours on repositories they can never see whole, and the industry's answer has been capacity. Context windows grew from 2K to 1M tokens in four years, and a large body of systems work optimizes the key--value (KV) cache, the per-token attention state that a long context creates inside the model (Section~\ref{sec:related} reviews it). All of this work makes the current turn cheaper. None of it gives an agent knowledge that survives the end of a session or a change of model.

This paper asks what should survive. Our central hypothesis: \textbf{coding agents need persistent, experience-aware repository memory to avoid repeatedly rediscovering knowledge across independent sessions.} We distinguish what the model can \emph{hold} (attention state over tokens it has read) from what the agent should \emph{remember}: the repository, its history, and its own experience with it. The latter, we propose, is a first-class software engineering artifact. Three measurements on real repositories ground this position.

\textbf{F1: Model-resident knowledge does not scale.} Django alone is 6.2M tokens, six times the largest production context window. No one proposes storing a repository as KV state; the measurement bounds the cost of any design that keeps knowledge inside the model. One billion tokens occupy about 4\,GB as indexed text, but about 300\,TB as fp16 KV state for a 70B-parameter model (Section~\ref{sec:study}). The KV form is also model-specific: it must be rebuilt whenever the model changes.

\textbf{F2: The directly connected code working set is small.} Across 752 bug-fix commits in Django, Flask, and scikit-learn, the median fix edits two files. These files, plus every file they import or are imported by (their one-hop import closure), amount to 2.7--9.9\% of the repository's code. This closure is only a proxy for the context a fix needs, generous in breadth but blind to configuration, runtime, and documentation dependencies. Even so, it suggests the code context of real maintenance fits in today's windows, \emph{if it can be found}.

\textbf{F3: Selection knowledge is not in the static structure.} In these commits, plus those of a fourth repository (852 in total), 76.5\% of the file pairs changed together by the same commit share no direct import edge. Extending to full transitive import reachability and test-naming links still leaves 47.3\% of the pairs disconnected. Yet co-edited pairs are $26\times$ more likely than random pairs to be import-linked: co-change is real structure, but static analysis sees only half of it. The other half must be mined from version history~\cite{zimmermann2005} and from experience, and it must persist.

Together, these findings reframe agent memory as two problems joined: a systems problem, a memory hierarchy with paging and eviction; and a software engineering problem, mining and maintaining the relationships that make a repository navigable. Section~\ref{sec:arch} sketches \sys{}, which organizes memory into three kinds: structural (parsed from code), historical (mined from version history), and experiential (written back by agents). The store lives with the repository and is exposed through nine \emph{page-safe} tools over the Model Context Protocol~\cite{mcp2024}: every response carries explicit pagination and budget state, so the agent, not a lossy heuristic, decides what enters its context. Section~\ref{sec:evidence} reports preliminary evidence for both the substrate and the mechanism, and Section~\ref{sec:plans} lays out the research plan.

\section{Motivating Study}
\label{sec:study}
We ran three measurements on six active open-source repositories (Django, Flask, Redis, requests, scikit-learn, Vue core), counting tokens with the \texttt{cl100k\_base} tokenizer. Scripts and raw results are in our replication package.\footnote{\url{https://doi.org/10.5281/zenodo.23615180}}

\subsection{E1: The Cost of Model-Resident Knowledge}
The six repositories hold 17.6M tokens of tracked source, documentation, and configuration in 70\,MB of text: Django 6.2M, scikit-learn 4.8M, Redis 4.7M, Vue core 1.5M, Flask 0.27M, and requests 0.15M. The measured text density is 4.17 bytes per token. One billion tokens of accumulated project knowledge cost about 4\,GB of SSD as text, or about 8\,GB with a search index. Held inside the model as fp16 KV state, the same tokens cost 119\,TB for an 8B-parameter model, 298\,TB for 70B, and 469\,TB for 405B (KV size per token is $2 \cdot L \cdot H_{kv} \cdot d \cdot 2$ bytes for $L$ layers, $H_{kv}$ KV heads, and head dimension $d$; the twos cover keys plus values at two bytes each, with grouped-query attention assumed). The gap is four to five orders of magnitude, and 2-bit quantization~\cite{kivi2024} narrows it by only one. This is a bound on representation cost, not a claim that anyone stores repositories as KV state: whatever long-lived agent memory is, it must live outside the model.

\subsection{E2: The Working Set of a Real Bug Fix}
We next measured how much of a repository a real bug fix touches. For the three Python-dominant repositories (Django, Flask, scikit-learn) we built the static import graph at the latest commit by parsing every file's import statements. We then sampled up to 300 recent non-merge commits per repository whose messages indicate a bug fix (the words \emph{fix}, \emph{bug}, or \emph{regression}) and that change at most 20 Python files, giving 752 commits. The median fix edits two files. Adding every file they import or are imported by gives the one-hop closure: 32 files (114,532 tokens, 2.7\% of the Python code) for Django, 24 files (4.6\%) for scikit-learn, and 2 files (9.9\%) for Flask. We claim only what this measures: the directly connected code around a fix is ten to forty times smaller than the repository. What a fix additionally needs, such as configuration, fixtures, runtime knowledge, and rationale, is exactly what Section~\ref{sec:arch} argues memory must supply.

\subsection{E3: What Static Structure Cannot See}
\label{sec:e3}
Finally, we asked how much of co-change is visible to static analysis. We took the same commits, plus those of requests, for a total of 852 commits and 3,695 pairs of files changed by the same commit. For each pair we checked progressively stronger static relationships. A direct import edge connects 23.5\% of the pairs. Full transitive reachability, in either direction, raises this to 52.0\%. Test-to-source name links raise it only to 52.7\%, leaving 47.3\% of the pairs disconnected under every static view we tried. At the same time, co-edited pairs are far from random: they are $26\times$ more likely than random file pairs to share a direct import edge (23.5\% versus 0.9\%; with transitive reachability, random pairs reach only 16.3\%). Files change together because of protocols, mirrored logic, and shared conventions, knowledge that lives in version history~\cite{zimmermann2005} and in maintainers' heads. A memory layer that indexes only parseable structure misses half of the real working set; it must learn edges from history and experience.

\section{\sys{}: Design}
\label{sec:arch}
\begin{figure*}[t]
\centering
\includegraphics[width=0.64\textwidth]{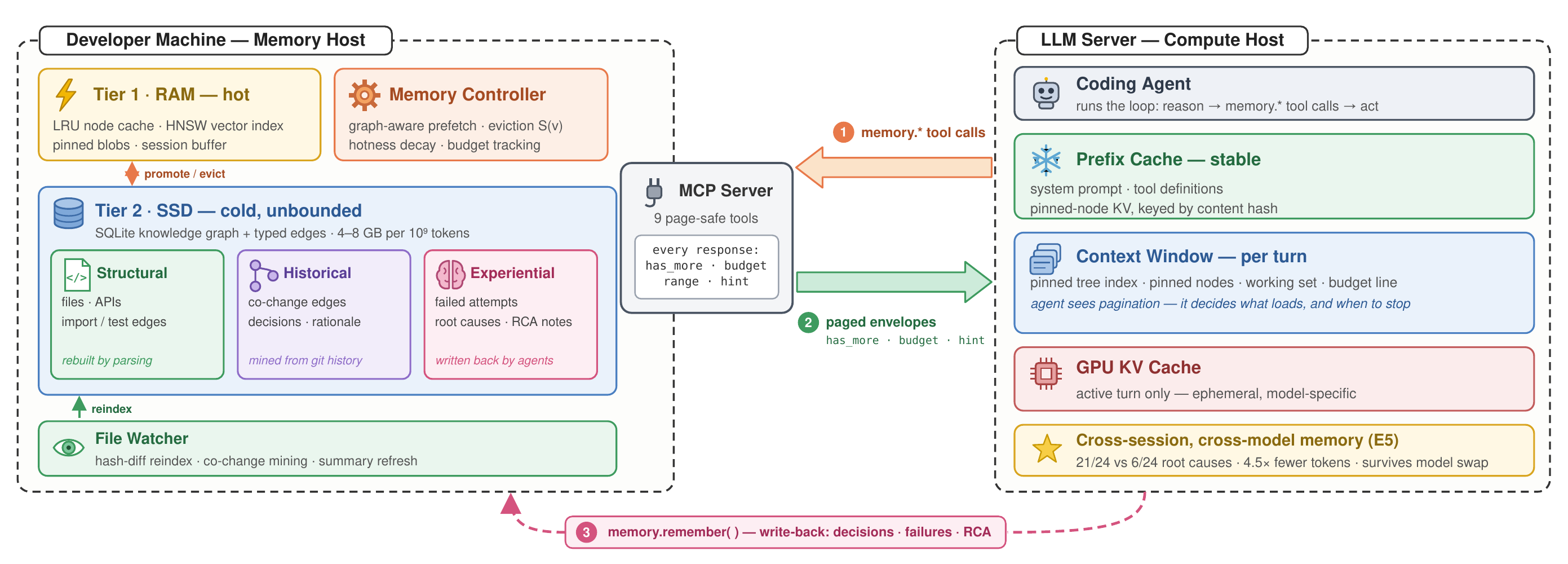}
\caption{\sys{} architecture. Knowledge lives as text and structure on the developer's machine; the model's KV cache holds only the active turn. The controller pages a scored working set across the MCP boundary under an explicit token budget.}
\label{fig:arch}
\end{figure*}

\sys{} is a memory hierarchy for coding agents (Fig.~\ref{fig:arch}). The analogy is the buffer pool of a database: a large, cheap store; a small, fast working set; and a controller that moves data between the two. The difference is where it lives: on the developer's machine, next to the repository, not inside the model or the serving stack. It holds structural memory (files, APIs, import and test edges; rebuilt by parsing), historical memory (co-change edges, decisions, design rationale; mined from version history), and experiential memory (failed attempts, root causes, debugging discoveries; written back by agents). E3 shows that structural memory alone recovers barely half of the working set. Historical and experiential memory are what existing retrieval stacks do not keep.

\subsection{Storage and Controller}
The cold tier is a SQLite graph. Its nodes are source files, tests, research notes, decisions, failures, and root-cause findings; its typed edges are \texttt{imports}, \texttt{tests}, \texttt{similar}, \texttt{sources}, and \texttt{co\_changes}, the last mined continuously from git history in direct response to E3. An HNSW index~\cite{hnsw2020} over locally computed embeddings serves similarity search. A file watcher re-indexes a file only when its content hash changes, and refreshes the co-change statistics. At 4 to 8\,GB per billion tokens, a decade of project memory fits on a laptop SSD.

The controller decides which nodes are \emph{resident}, that is, held in RAM and in the agent's context. It scores each node
\[
S(v) \;=\; w_1 G(v) + w_2 C(v) + w_3 H(v) + w_4 P(v),
\]
where $G$ is graph proximity to the files under edit, $C$ is co-change affinity, $H$ is time-decayed access frequency, and $P$ is explicit pin priority. We present this linear form as a research hypothesis, not a tuned policy; RQ3 (Section~\ref{sec:plans}) asks whether any semantic scoring beats recency-only baselines. Eviction follows the score downward: pinned nodes are never evicted; direct neighbors of the files under edit keep interface-level detail; other warm nodes degrade to one-line summaries; everything else returns to SSD, less than 3\,ms away (Section~\ref{sec:evidence}). When free context drops below 15\%, the controller evicts the largest and coldest nodes first and reports what it removed.

\subsection{Nine Page-Safe Tools}
The agent reaches the store through nine MCP tools. Five read: \texttt{tree} lists the store as a hierarchy; \texttt{search} combines embedding similarity, keyword match, and recency; \texttt{get} returns a node at a chosen detail level (\texttt{summary}, \texttt{api}, or \texttt{full}) in fixed-size character pages; \texttt{walk} follows typed edges; \texttt{budget} reports token usage. One writes: \texttt{remember} persists a new knowledge node. Three control residency: \texttt{pin}/\texttt{unpin} holds a node in context, \texttt{prefetch} warms nodes ahead of need, and \texttt{evict} releases them. Every response is one envelope carrying the content, the pagination state (\texttt{page}, \texttt{total\_pages}, \texttt{has\_more}), the remaining token budget, and a hint such as ``Page 2/3; \texttt{create()} is on this page.''

We call this contract \emph{page safety}: no tool ever silently truncates. The agent always knows whether more content exists, which range it holds, and what the call cost, so it can decide that page 2 of 3 already contains the answer and stop. Write-back closes the loop. At the end of a task the agent persists what it learned, and the next session, possibly run by a different model, inherits it.

\section{Design Rationale}
\label{sec:rationale}
Five decisions define \sys{}. Each follows from a finding or from a measured cost.

\textbf{R1. Store text and structure, not tensors.} KV state is model-specific and holds only tokens some model has already read; text plus a graph is $10^4$ to $10^5$ times smaller (F1) and holds knowledge no model has read yet. Storage portability is free: text survives any model swap. Retrieval portability is not: embeddings must be recomputed for each encoder, and ranking behavior may shift. The KV-cache work reviewed in Section~\ref{sec:related} is complementary, not competing: it makes the current turn cheaper, and \sys{} decides what the turn contains.

\textbf{R2. Make paging legible to the agent.} Retrieval pipelines truncate silently, and models demonstrably misuse what is placed into long contexts~\cite{lostmiddle2024}. An agent can manage its memory only if it can see that memory's boundaries. The explicit \texttt{has\_more} and budget fields turn context into a resource the agent reasons about, and F2 makes this practical: the working set is small enough to hold, and reason about, in one window.

\textbf{R3. Learn the edges; keep the failures.} E3 shows that even transitive static structure misses half of what changes together, so \sys{} mines co-change edges from version history~\cite{zimmermann2005} and lets agents write \texttt{failure} and \texttt{rca} (root-cause analysis) nodes linked to the code they concern. We conjecture that this negative knowledge, the record of what was tried and why it failed, is the store's highest-value content. Nothing prevents a RAG pipeline from storing failure notes, but we are not aware of deployed coding-agent stacks that keep them as durable, linkable records, yet such records are exactly what stop an agent from spending its budget rediscovering a dead end (RQ4, Section~\ref{sec:plans}).

\textbf{R4. Semantic signals for an OS-style policy.} Classic cache-replacement policies treat pages as opaque; \sys{} instead sees graph distance, co-change affinity, and pin intent. Whether these signals beat a least-recently-used (LRU) policy is an open, testable question (RQ3), not an assumption.

\textbf{R5. Live with the code; compose with serving caches.} Hosting memory on the developer's machine gives freshness (the watcher sees every save) and limits data exposure. The repository as a whole never leaves the machine. Only the scoped working set is sent to a hosted model, and a locally run model avoids even that. Pinned nodes enter the context as stable blocks keyed by content hash, and serving-side prefix caches, which reuse the attention state of an unchanged prefix, skip recomputing them~\cite{cachedattention2024,mooncake2024,cacheblend2024}.

\section{Preliminary Evidence}
\label{sec:evidence}
\subsection{E4: Storage and Retrieval Feasibility}
We built the cold tier at realistic scale on a consumer laptop (Core Ultra~9, 32\,GB RAM, NVMe SSD, CPU only): $10^6$ nodes holding about 440M tokens of synthetic content, $5{\times}10^6$ edges, and 384-dimensional vectors; the store occupies 3.7\,GB and builds in under four minutes. Across 200 end-to-end retrievals (a vector search with $k{=}20$, a one-hop graph walk, and loading the 20 result nodes) the median took 1.70\,ms and the 99th percentile 2.76\,ms; two-hop walks took 0.95\,ms at the median. We claim exactly two things: the store fits on commodity hardware, and its basic operations are orders of magnitude faster than an LLM turn. Retrieval quality on real embeddings, controller overhead, and the cost of keeping the index fresh as files change remain unmeasured (Section~\ref{sec:plans}).

\subsection{E5: Does Inherited Memory Prevent Repeated Work?}
\label{sec:e5}
We implemented a prototype of \sys{}'s storage and tool layer, a SQLite store with lexical full-text retrieval and four of the nine tools (\texttt{tree}, \texttt{search}, \texttt{get}, \texttt{remember}) with the full page-safe envelope, and ran the multi-session design our hypothesis implies. The tasks are eight real bug-fix commits from requests and Flask. Each task gives the agent a behaviorally described symptom, with file paths removed, and the repository tree from before the fix, so the bug is present; ground truth is the source files the human fix changed, and success means naming the root-cause file (a conservative, file-level criterion). In Session~1, a Claude Sonnet agent (the \emph{writer}) investigates the bug and persists \texttt{rca} and \texttt{failure} notes. After a full context reset, a different, smaller model (Claude Haiku, the \emph{reader}) diagnoses the same symptom in Session~2 under identical prompts. Condition~A strips the notes; the reader keeps full lexical search over the repository, a retrieval-only baseline in the style of retrieval-augmented generation (RAG). Condition~B inherits the notes, the same retrieval plus experiential memory; A and B ran three trials per task. Condition~P (one trial) inherits the pooled 26 notes of all eight investigations (mostly unrelated bugs, some from the other project) to test retrieval as memory accumulates. Condition~C (two trials) places the entire repository in the context window in a single call, with no tools, possible only because requests and Flask are our census's two smallest repositories. 

\begin{table}[t]
\caption{E5: Session-2 outcomes across 8 real bugs (medians). Notes were written by a different model than the one using them. In-tok for C is per call.}
\label{tab:agent}
\centering
\footnotesize
\setlength{\tabcolsep}{4.5pt}
\begin{tabular}{@{}lrrrr@{}}
\toprule
Condition & Success & Calls & Reads & In-tok \\
\midrule
A: retrieval only      & \evSucA{} & \evCallsA{} & \evGetsA{} & \evTokA{}K \\
B: $+$ experiential memory & \evSucB{} & \evCallsB{} & \evGetsB{} & \evTokB{}K \\
P: B, pooled store       & 8/8 & 8 & 4.5 & 41.7K \\
C: whole repo in context & 14/16 & --- & --- & 164.4K \\
\bottomrule
\end{tabular}
\end{table}

Table~\ref{tab:agent} summarizes the outcomes. With inherited notes (B), the reader succeeded in 21 of 24 trials at $4.5\times$ lower median input-token cost. The retrieval-only baseline (A) succeeded in 6 of 24 and typically exhausted its 16-round tool budget; a one-sided Fisher exact test gives $p=1.3{\times}10^{-5}$. Accumulation did not hurt: with all 26 notes pooled (P), success was 8 of 8 at the same cost, including the one task (an IPv6 bug) that had failed in two of three B trials. The writer solved all eight tasks (median 83K tokens, paid once), so the probe measures transfer rather than writer errors. The full-context condition (C) succeeded in 14 of 16 trials, exactly B's 87.5\% success rate, but a single call cost $3.8\times$ the tokens of an entire memory session (164K versus 43K), the cost recurs on every further call, nothing persists, and the approach is unavailable for a repository like Django, which is $31\times$ the reader's 200K-token window. The one task C failed in both trials, memory solved in all three. This probe is small (eight tasks, two repositories, re-diagnosis rather than transfer to a new task), but it exercises the full mechanism: knowledge one model wrote through the page-safe interface changed what a different model did with a cold context, which no window growth provides.

\section{Related Work}
\label{sec:related}
KV-cache management spans quantization~\cite{kivi2024}, token selection~\cite{h2o2023,snapkv2024,quest2024,magicpig2024,streamingllm2023}, paged serving memory~\cite{pagedattention2023}, multi-tier offloading~\cite{flexgen2023,infinigen2024,kvswap2025,kvdrive2026,shadowkv2024}, cross-request reuse~\cite{cachegen2023,cachedattention2024,mooncake2024,cacheblend2024,lmcache2025}, and retrieval over the cache itself~\cite{retrievalattention2024,freekv2025,pariskv2026,infllm2024,memorizing2022}, with surveys in~\cite{kvsurvey2024,scbench2024}. All of this work manages model-specific state for tokens already read within one serving context; none of it survives a session or model boundary. \sys{} sits above this layer and composes with it (R5).

Our differentiation from the closest neighbors is the combination of three properties. First, \sys{} is repository-native: a typed code graph, co-change mining, and a file watcher. MemGPT~\cite{memgpt2023}, the closest system in spirit, pages conversation history and knows nothing about code. Second, \sys{} is experience-aware: failures and root causes are durable nodes linked to the code they concern. Repository RAG~\cite{rag2020} and code knowledge graphs index what the code \emph{is}, not what was \emph{learned} about it. Third, \sys{} is budget-controlled: the page-safe contract makes truncation and cost visible to the agent. Conventional retrieval interfaces return the $k$ highest-ranked results and discard the rest silently. Co-change mining itself is a classic technique from mining software repositories research~\cite{zimmermann2005}; \sys{} turns it into live edges traversed during a task.

\section{Future Plans}
\label{sec:plans}
E5 shows the mechanism works at small scale; the program now is to test it at a scale that can serve as evidence.
\emph{RQ1:} Does persistent repository memory improve multi-session task success? We will use repeated and related SWE-bench~\cite{swebench2024} instances, measuring success, re-reads, repeated failures, and turns.
\emph{RQ2:} Do historical and experiential edges retrieve the files a fix actually needs better than static structure alone? Ground truth: the files touched by real fixes, as in E2 and E3.
\emph{RQ3:} Does semantic eviction ($S(v)$, Section~\ref{sec:arch}) beat LRU under a fixed context budget?
\emph{RQ4:} Does negative knowledge reduce repeated failed approaches? We will remove the \texttt{failure} and \texttt{rca} nodes and compare.
\emph{RQ5:} Does memory transfer across models and frameworks, generalizing E5?
Beyond these lie memory hygiene (a wrong note poisons future sessions; written-back web content is a prompt-injection channel), summary faithfulness, and team-shared stores.

\emph{Limitations.} E2 and E3 analyze Python files only, and their import graph is built at the latest commit while the commits analyzed are older, so files may have moved or changed in between. Message-keyword commit selection may bias the sample toward well-described fixes. The one-hop closure is a proxy for required context; RQ2 validates it against files developers actually consulted. E4 uses synthetic content, and in E5 we did not control the quality of the writer's notes, the retrieval baseline is lexical rather than embedding-based, and we did not test plain session-summary carry-over (RQ1).

\section{Conclusion}
The context window is a cache; the field is making it larger and cheaper, but the harder problem is what it should contain. F1 to F3 showed that the knowledge an agent needs is small, half-invisible to static analysis, and affordable only as text beside the code. \sys{} sketches this layer, E5 shows it working across a model boundary, and the next step is evidence at scale.

\balance
\bibliographystyle{IEEEtran}
\bibliography{refs}

@inproceedings{streamingllm2023,
  title        = {Efficient Streaming Language Models with Attention Sinks},
  author       = {Guangxuan Xiao and Yuandong Tian and Beidi Chen and Song Han and Mike Lewis},
  booktitle    = {Proc.\ International Conference on Learning Representations (ICLR)},
  year         = {2024},
  eprint       = {2309.17453},
  archivePrefix= {arXiv}
}

@article{memgpt2023,
  title        = {MemGPT: Towards LLMs as Operating Systems},
  author       = {Charles Packer and Sarah Wooders and Kevin Lin and Vivian Fang and Shishir G. Patil and Ion Stoica and Joseph E. Gonzalez},
  journal      = {arXiv preprint arXiv:2310.08560},
  year         = {2023},
  eprint       = {2310.08560},
  archivePrefix= {arXiv}
}

@article{cachegen2023,
  title        = {CacheGen: KV Cache Compression and Streaming for Fast Large Language Model Serving},
  author       = {Yuhan Liu and Hanchen Li and Yihua Cheng and Siddhant Ray and Yuyang Huang and Qizheng Zhang and Kuntai Du and Jiayi Yao and Shan Lu and Ganesh Ananthanarayanan and Michael Maire and Henry Hoffmann and Ari Holtzman and Junchen Jiang},
  journal      = {arXiv preprint arXiv:2310.07240},
  year         = {2023},
  eprint       = {2310.07240},
  archivePrefix= {arXiv}
}

@inproceedings{memorizing2022,
  title        = {Memorizing Transformers},
  author       = {Yuhuai Wu and Markus N. Rabe and DeLesley Hutchins and Christian Szegedy},
  booktitle    = {Proc.\ International Conference on Learning Representations (ICLR)},
  year         = {2022},
  eprint       = {2203.08913},
  archivePrefix= {arXiv}
}

@inproceedings{h2o2023,
  title        = {H$_2$O: Heavy-Hitter Oracle for Efficient Generative Inference of Large Language Models},
  author       = {Zhenyu Zhang and Ying Sheng and Tianyi Zhou and Tianlong Chen and Lianmin Zheng and Ruisi Cai and Zhao Song and Yuandong Tian and Christopher R\'e and Clark Barrett and Zhangyang Wang and Beidi Chen},
  booktitle    = {Advances in Neural Information Processing Systems (NeurIPS)},
  volume       = {36},
  year         = {2023},
  eprint       = {2306.14048},
  archivePrefix= {arXiv}
}

@inproceedings{pagedattention2023,
  title        = {Efficient Memory Management for Large Language Model Serving with PagedAttention},
  author       = {Woosuk Kwon and Zhuohan Li and Siyuan Zhuang and Ying Sheng and Lianmin Zheng and Cody Hao Yu and Joseph E. Gonzalez and Hao Zhang and Ion Stoica},
  booktitle    = {Proc.\ 29th ACM Symposium on Operating Systems Principles (SOSP)},
  year         = {2023},
  doi          = {10.1145/3600006.3613165},
  eprint       = {2309.06180},
  archivePrefix= {arXiv}
}

@article{infllm2024,
  title        = {InfLLM: Training-Free Long-Context Extrapolation for LLMs with an Efficient Context Memory},
  author       = {Chaojun Xiao and Pengle Zhang and Xu Han and Guangxuan Xiao and Yankai Lin and Zhengyan Zhang and Zhiyuan Liu and Maosong Sun},
  journal      = {arXiv preprint arXiv:2402.04617},
  year         = {2024},
  eprint       = {2402.04617},
  archivePrefix= {arXiv}
}

@article{kivi2024,
  title        = {KIVI: A Tuning-Free Asymmetric 2bit Quantization for KV Cache},
  author       = {Zirui Liu and Jiayi Yuan and Hongye Jin and Shaochen Zhong and Zhaozhuo Xu and Vladimir Braverman and Beidi Chen and Xia Hu},
  journal      = {arXiv preprint arXiv:2402.02750},
  year         = {2024},
  eprint       = {2402.02750},
  archivePrefix= {arXiv}
}

@article{flexgen2023,
  title        = {FlexGen: High-Throughput Generative Inference of Large Language Models with a Single GPU},
  author       = {Ying Sheng and Lianmin Zheng and Binhang Yuan and Zhuohan Li and Max Ryabinin and Daniel Y. Fu and Zhiqiang Xie and Beidi Chen and Clark Barrett and Joseph E. Gonzalez and Percy Liang and Christopher Ré and Ion Stoica and Ce Zhang},
  journal      = {arXiv preprint arXiv:2303.06865},
  year         = {2023},
  eprint       = {2303.06865},
  archivePrefix= {arXiv}
}

@article{cachedattention2024,
  title        = {Cost-Efficient Large Language Model Serving for Multi-turn Conversations with CachedAttention},
  author       = {Bin Gao and Zhuomin He and Puru Sharma and Qingxuan Kang and Djordje Jevdjic and Junbo Deng and Xingkun Yang and Zhou Yu and Pengfei Zuo},
  journal      = {arXiv preprint arXiv:2403.19708},
  year         = {2024},
  eprint       = {2403.19708},
  archivePrefix= {arXiv}
}

@inproceedings{scbench2024,
  title        = {SCBench: A KV Cache-Centric Analysis of Long-Context Methods},
  author       = {Yucheng Li and Huiqiang Jiang and Qianhui Wu and Xufang Luo and Surin Ahn and Chengruidong Zhang and Amir H. Abdi and Dongsheng Li and Jianfeng Gao and Yuqing Yang and Lili Qiu},
  booktitle    = {Proc.\ International Conference on Learning Representations (ICLR)},
  year         = {2025},
  eprint       = {2412.10319},
  archivePrefix= {arXiv}
}

@article{shadowkv2024,
  title        = {ShadowKV: KV Cache in Shadows for High-Throughput Long-Context LLM Inference},
  author       = {Hanshi Sun and Li-Wen Chang and Wenlei Bao and Size Zheng and Ningxin Zheng and Xin Liu and Harry Dong and Yuejie Chi and Beidi Chen},
  journal      = {arXiv preprint arXiv:2410.21465},
  year         = {2024},
  eprint       = {2410.21465},
  archivePrefix= {arXiv}
}

@article{mooncake2024,
  title        = {Mooncake: A KVCache-centric Disaggregated Architecture for LLM Serving},
  author       = {Ruoyu Qin and Zheming Li and Weiran He and Mingxing Zhang and Yongwei Wu and Weimin Zheng and Xinran Xu},
  journal      = {arXiv preprint arXiv:2407.00079},
  year         = {2024},
  eprint       = {2407.00079},
  archivePrefix= {arXiv}
}

@article{cacheblend2024,
  title        = {CacheBlend: Fast Large Language Model Serving for RAG with Cached Knowledge Fusion},
  author       = {Jiayi Yao and Hanchen Li and Yuhan Liu and Siddhant Ray and Yihua Cheng and Qizheng Zhang and Kuntai Du and Shan Lu and Junchen Jiang},
  journal      = {arXiv preprint arXiv:2405.16444},
  year         = {2024},
  eprint       = {2405.16444},
  archivePrefix= {arXiv}
}

@article{infinigen2024,
  title        = {InfiniGen: Efficient Generative Inference of Large Language Models with Dynamic KV Cache Management},
  author       = {Wonbeom Lee and Jungi Lee and Junghwan Seo and Jaewoong Sim},
  journal      = {arXiv preprint arXiv:2406.19707},
  year         = {2024},
  eprint       = {2406.19707},
  archivePrefix= {arXiv}
}

@article{quest2024,
  title        = {Quest: Query-Aware Sparsity for Efficient Long-Context LLM Inference},
  author       = {Jiaming Tang and Yilong Zhao and Kan Zhu and Guangxuan Xiao and Baris Kasikci and Song Han},
  journal      = {arXiv preprint arXiv:2406.10774},
  year         = {2024},
  eprint       = {2406.10774},
  archivePrefix= {arXiv}
}

@article{magicpig2024,
  title        = {MagicPIG: LSH Sampling for Efficient LLM Generation},
  author       = {Zhuoming Chen and Ranajoy Sadhukhan and Zihao Ye and Yang Zhou and Jianyu Zhang and Niklas Nolte and Yuandong Tian and Matthijs Douze and Leon Bottou and Zhihao Jia and Beidi Chen},
  journal      = {arXiv preprint arXiv:2410.16179},
  year         = {2024},
  eprint       = {2410.16179},
  archivePrefix= {arXiv}
}

@article{snapkv2024,
  title        = {SnapKV: LLM Knows What You are Looking for Before Generation},
  author       = {Yuhong Li and Yingbing Huang and Bowen Yang and Bharat Venkitesh and Acyr Locatelli and Hanchen Ye and Tianle Cai and Patrick Lewis and Deming Chen},
  journal      = {arXiv preprint arXiv:2404.14469},
  year         = {2024},
  eprint       = {2404.14469},
  archivePrefix= {arXiv}
}

@article{retrievalattention2024,
  title        = {RetrievalAttention: Accelerating Long-Context LLM Inference via Vector Retrieval},
  author       = {Di Liu and Meng Chen and Baotong Lu and Huiqiang Jiang and Zhenhua Han and Qianxi Zhang and Qi Chen and Chengruidong Zhang and Bailu Ding and Kai Zhang and Chen Chen and Fan Yang and Yuqing Yang and Lili Qiu},
  journal      = {arXiv preprint arXiv:2409.10516},
  year         = {2024},
  eprint       = {2409.10516},
  archivePrefix= {arXiv}
}

@article{lmcache2025,
  title        = {LMCache: An Efficient KV Cache Layer for Enterprise-Scale LLM Inference},
  author       = {Yuhan Liu and Yihua Cheng and Jiayi Yao and Yuwei An and Xiaokun Chen and Shaoting Feng and Yuyang Huang and Samuel Shen and Rui Zhang and Kuntai Du and Junchen Jiang},
  journal      = {arXiv preprint arXiv:2510.09665},
  year         = {2025},
  eprint       = {2510.09665},
  archivePrefix= {arXiv}
}

@inproceedings{freekv2025,
  title        = {FreeKV: Boosting KV Cache Retrieval for Efficient LLM Inference},
  author       = {Guangda Liu and Chengwei Li and Zhenyu Ning and Jing Lin and Yiwu Yao and Danning Ke and Minyi Guo and Jieru Zhao},
  booktitle    = {Proc.\ International Conference on Learning Representations (ICLR)},
  year         = {2026},
  eprint       = {2505.13109},
  archivePrefix= {arXiv}
}

@article{kvsurvey2024,
  title        = {A Survey on Large Language Model Acceleration based on KV Cache Management},
  author       = {Haoyang Li and Yiming Li and Anxin Tian and Tianhao Tang and Zhanchao Xu and Xuejia Chen and Nicole Hu and Wei Dong and Qing Li and Lei Chen},
  journal      = {arXiv preprint arXiv:2412.19442},
  year         = {2024},
  eprint       = {2412.19442},
  archivePrefix= {arXiv}
}

@article{kvswap2025,
  title        = {KVSwap: Disk-aware KV Cache Offloading for Long-Context On-device Inference},
  author       = {Huawei Zhang and Chunwei Xia and Zheng Wang},
  journal      = {arXiv preprint arXiv:2511.11907},
  year         = {2025},
  eprint       = {2511.11907},
  archivePrefix= {arXiv}
}

@article{kvdrive2026,
  title        = {KVDrive: A Holistic Multi-Tier KV Cache Management System for Long-Context LLM Inference},
  author       = {Jian Lin and Jiazhi Mi and Zicong Hong and Haodong Wang and Qianli Liu and Haodyue Zhang and Peng Li and Song Guo},
  journal      = {arXiv preprint arXiv:2605.18071},
  year         = {2026},
  eprint       = {2605.18071},
  archivePrefix= {arXiv}
}

@article{pariskv2026,
  title        = {ParisKV: Fast and Drift-Robust KV-Cache Retrieval for Long-Context LLMs},
  author       = {Yanlin Qi and Xinhang Chen and Huiqiang Jiang and Qitong Wang and Botao Peng and Themis Palpanas},
  journal      = {arXiv preprint arXiv:2602.07721},
  year         = {2026},
  eprint       = {2602.07721},
  archivePrefix= {arXiv}
}

@inproceedings{swebench2024,
  title     = {{SWE}-bench: Can Language Models Resolve Real-World {GitHub} Issues?},
  author    = {Carlos E. Jimenez and John Yang and Alexander Wettig and Shunyu Yao and Kexin Pei and Ofir Press and Karthik Narasimhan},
  booktitle = {Proc.\ ICLR},
  year      = {2024}
}

@article{lostmiddle2024,
  title   = {Lost in the Middle: How Language Models Use Long Contexts},
  author  = {Nelson F. Liu and Kevin Lin and John Hewitt and Ashwin Paranjape and Michele Bevilacqua and Fabio Petroni and Percy Liang},
  journal = {Transactions of the Association for Computational Linguistics},
  volume  = {12},
  pages   = {157--173},
  year    = {2024},
  doi     = {10.1162/tacl\_a\_00638}
}

@article{zimmermann2005,
  title   = {Mining Version Histories to Guide Software Changes},
  author  = {Thomas Zimmermann and Peter Wei{\ss}gerber and Stephan Diehl and Andreas Zeller},
  journal = {IEEE Transactions on Software Engineering},
  volume  = {31},
  number  = {6},
  pages   = {429--445},
  year    = {2005}
}

@article{hnsw2020,
  title   = {Efficient and Robust Approximate Nearest Neighbor Search Using Hierarchical Navigable Small World Graphs},
  author  = {Yu A. Malkov and Dmitry A. Yashunin},
  journal = {IEEE Transactions on Pattern Analysis and Machine Intelligence},
  volume  = {42},
  number  = {4},
  pages   = {824--836},
  year    = {2020}
}

@inproceedings{rag2020,
  title     = {Retrieval-Augmented Generation for Knowledge-Intensive {NLP} Tasks},
  author    = {Patrick Lewis and Ethan Perez and Aleksandra Piktus and Fabio Petroni and Vladimir Karpukhin and Naman Goyal and Heinrich K{\"u}ttler and Mike Lewis and Wen-tau Yih and Tim Rockt{\"a}schel and Sebastian Riedel and Douwe Kiela},
  booktitle = {Proc.\ NeurIPS},
  year      = {2020}
}

@misc{mcp2024,
  title        = {Model Context Protocol},
  author       = {{Anthropic}},
  year         = {2024},
  howpublished = {\url{https://modelcontextprotocol.io}}
}

\end{document}